\documentstyle[12pt]{article}
\begin{document}
\begin{center}

\author{M. N. Vinoj$^1$ and V. C. Kuriakose$^2$ \\
Department of Physics\\
Cochin University of Science and Techology\\
Kochi-682 022, India.}
\title{Multisoliton solutions and integrability aspects of coupled higher-order
nonlinear Schr\"odinger equations.}
\date{}
\maketitle
\end{center}
\begin{abstract}
Using Painlev\'e singularity structure analysis, we show that coupled
higher-order nonlinear Schr\"odinger (CHNLS) equations admit Painlev\'e
property. Using the results of Painlev\'e analysis, we succeed in Hirota
bilinearizing the CHNLS equations for the integrable cases. Solving the
Hirota bilinear equations, one soliton and two soliton solutions are
explicitly obtained. Lax pairs are explicitly constructed.\vspace{2cm}\\PACS
number(s): 42.81. Dp, 02.30 Jr, 04.30 Nk\vspace{2cm}\\1) Electronic address:
vinojmn@cusat.ac.in\\2) Electronic address: vck@cusat.ac.in\vspace{1.5cm}\\(to appear in 
Phys. Rev.E)\newpage\ 
\end{abstract}

\begin{center}
{\bf I. INTRODUCTION}
\end{center}

The last three decades witnessed extensive theoretical and experimental
studies on optical solitons because of their potential applications in long
distance communication. The invention of high-intensity lasers helped
Mollenauer {\it et al} [1] to verify experimentally the pioneering
theoretical work on optical solitons initiated by Hasegawa and Tappert [2].
The solitons, localised-in-time optical pulses, evolve from a nonlinear
change in the refractive index of the material, known as Kerr effect,
induced by the light intensity-distribution. When the combined effect of the
intensity-dependent refractive index nonlinearity and the
frequency-dependent pulse dispersion exactly compensate each other, the
pulse propagates without any change in its shape, being self-trapped by the
waveguide nonlinearity. The propagation of optical solitons in a nonlinear
dispersive optical fibre is governed by the well-known nonlinear
Schr\"odinger (NLS) equation of the form:

\begin{equation}
iu_t+\alpha \;u_{zz}+\beta \left| u\right| ^2\;u=0 
\end{equation}
\\where u is the complex amplitude of the pulse envelope, $\alpha $ and $
\beta $ are the group velocity dispersion (GVD) and self-phase modulation
parameters respectively and subscripts z and t represent the spatial and
temporal coordinates respectively.

When ultrashort pulses (USPs) are transmitted through fibres, higher-order
effects such as third-order dispersion (TOD), Kerr dispersion and stimulated
Raman scattering (SRS) come into play as experimentally reported by Mitschke
and Mollenauer[3]. The Kerr dispersion, also known as self-steepening, is
caused by the intensity dependence of the group velocity which results in
asymmetrical spectral broadening of the pulse since the peak of the pulse
travels slower than the wings. The SRS causes a self-frequency shift which
is a self-induced red shift in the pulse spectrum as the low-frequency
components of the pulse obtain Raman gain at the expense of the
high-frequency components. With the inclusion of all these effects, Kodama
and Hasegawa [4] have proposed that the dynamics of femtosecond pulse
propagation be governed by a higher-order NLS (HNLS) equation. The HNLS
equation allows soliton-type propagation only for certain choices of
parameters [5].

A coupled NLS equation was proposed by Manakov, by taking into account the
fact that the total field comprises two fields with left and right
polarizations [7]. The coupled equation takes the form:

\begin{equation}
\begin{array}{c}
iu_t+c_1u_{zz}+\left( \alpha 
\left| u\right| ^2+\beta \left| v\right| ^2\right) u=0 \\  \\ 
iv_t+c_2v_{zz}+\left( \beta \left| u\right| ^2+\gamma \left|
v\right| ^2\right) v=0 
\end{array}
\end{equation}

The above equations are integrable only for the following parametric
choices: (i) $c_1=c_2,$ $\alpha =\beta =\gamma $ and (ii) $c_1=-c_2,$ $
\alpha =\beta =\gamma .$ Recently, for USPs, Eq.(2) was generalized to a set
of coupled higher-order NLS (CHNLS)equation which can be derived from the
Maxwell's equations in order to investigate the effects of birefringence on
pulse propagation in femtosecond regime [8, 9]. The general form of CHNLS
equations are:

\begin{equation}
\begin{array}{c}
iu_t+u_{zz}+2\left( \left| u\right| ^2+\left| v\right| ^2\right) u \\ 
- 
i\lambda \left[ \beta _1u_{zzz}+\beta _2\left( \left|
u\right| ^2+\left| v\right| ^2\right) u_z+\beta _3\left( \left| u\right|
^2+\left| v\right| ^2\right) _zu\right] =0 \\  \\ 
iv_t+v_{zz}+2\left( \left| u\right| ^2+\left| v\right| ^2\right) v \\ 
-i\lambda \left[ \beta _{1}v_{zzz}+\beta _2\left( \left|
u\right| ^2+\left| v\right| ^2\right) v_z+\beta _3\left( \left| u\right|
^2+\left| v\right| ^2\right) _zv\right] =0 
\end{array}
\end{equation}

In general, the above equations are not completely integrable. However, if
some restrictions are imposed on the parametric values, one can obtain
several integrable, soliton- possessing NLS-type equations: (i) $\lambda =0,$
NLS; (ii) $\beta _1:$ $\beta _2:$ $\beta _3=0:1:1,$ derivative NLS [10];
(iii) $\beta _1:$ $\beta _2:$ $\beta _3=0:1:0$, derivative mixed NLS [10];
(iv) $\beta _1:$ $\beta _2:$ $\beta _3=1:6:0,$ the Hirota equation [11] and
(v) $\beta _1:$ $\beta _2:$ $\beta _3=1:6:3,$ Sasa-Satsuma equation [8].
Along the lines of Refs.[12, 13] we choose $\beta _1=1,$ $\beta _2=6,$ $%
\beta _3=3.$ For this choice of parameters, Eqs.(3) become 
\begin{equation}
\begin{array}{c}
iu_t+u_{zz}+2\left( \left| u\right| ^2+\left| v\right| ^2\right) u \\ 
- 
i\lambda \left[ u_{zzz}+6\left( \left| u\right| ^2+\left| v\right|
^2\right) u_z+3\left( \left| u\right| ^2+\left| v\right| ^2\right)
_zu\right] =0 \\  \\ 
iv_t+v_{zz}+2\left( \left| u\right| ^2+\left| v\right| ^2\right) v \\ 
-i\lambda \left[ v_{zzz}+6\left( \left| u\right| ^2+\left| v\right|
^2\right) v_z+3\left( \left| u\right| ^2+\left| v\right| ^2\right)
_zv\right] =0 
\end{array}
\end{equation}

The plan of this paper is as follows. In Sec. II, we establish the
Painlev\'e property of the above system of equations. In Sec. III, we
rewrite Eqs. (7) in a Hirota bilinear form. Section IV is devoted to the
construction of exact one soliton and two soliton solutions. In Section V,
Lax pair for CHNLS equations are obtained. Conclusion and discussion of the
present calculation are presented in Section VI.

\begin{center}
{\bf II. PAINLEV\'E ANALYSIS OF CHNLS EQUATIONS}
\end{center}

In this section we study Painlev\'e analysis of Eqs. (4). The motivation
behind this exercise is that P condition is necessary one for studying the
integrablity of nonlinear partial differential equations [14-19] and helps
construct solutions. The method for applying Painlev\'e test to partial
differential equations were introduced by Weiss, Tabor, and Carnavale [15]
with simplification due to Kruskal [16] involves seeking a solution of a
given partial differential equation in the form:

\begin{equation}
\begin{array}{c}
u(z,t)=\phi ^\alpha \sum_{j=0}^\infty u_j(t)\phi ^j(z,t), 
u_0\neq 0 \\  \\ 
v(z,t)=\phi ^\beta \sum_{j=0}^\infty v_j(t)\phi ^j(z,t),v_0\neq 0 
\end{array}
\end{equation}
\\with

\begin{equation}
\phi (z,t)=z+\psi (t)=0 
\end{equation}
\\where $\psi (t)$ is an arbitrary analytic function of t, $u_j(t)$ and $
v_j(t)$ $($ $j=0,1,2,....,)$, in the neighbourhood of a noncharacteristic
movable singularity manifold defined by $\phi =0.$

Apart from providing the integrability property of a given nonlinear partial
differential equations, P analysis also provides information about
B\"acklund transformation(BT), Lax pair, Hirota's bilinear representation,
special and rational solutions, etc.[15-17]. Many of these results are
obtained by truncating the Laurent series at a constant level term [18, 19].

In order to investigate the integrability properties of Eqs.(7), we rewrite
it in terms of four complex functions a, b, c and d by defining $u=a,$ $
u^{*}=b,$ $v=c,$ $v^{*}=d$. Consequently, we have the following equations:

\begin{equation}
\begin{array}{c}
ia_t+a_{zz}+2(ab+cd)a-i\lambda [a_{zzz}+6(ab+cd)a_z+3(ab+cd)_za]=0 \\  
\\ 
-ib_t+b_{zz}+2(ab+cd)b+i\lambda [b_{zzz}+6(ab+cd)b_z+3(ab+cd)_zb]=0 \\  
\\ 
ic_t+c_{zz}+2(ab+cd)c-i\lambda [c_{zzz}+6(ab+cd)c_z+3(ab+cd)_zc]=0 \\  
\\ 
-id_t+d_{zz}+2(ab+cd)d+i\lambda [d_{zzz}+6(ab+cd)d_z+3(ab+cd)_zd]=0 
\end{array}
\end{equation}

The Painlev\'e analysis essentially consists of four main stages (i)
determination of leading-order behaviour, (ii) identifying the resonance
values, (iii) verifying that at resonance values sufficient number of
arbitrary functions exist without the introduction of movable critical
manifold, and (iv) identifies connection with the integrability properties
like Lax pair, BT.

Looking at the leading order behaviour, we substitute $a\simeq a_0\phi
^{\alpha _1},$ $b\simeq b_0\phi ^{\alpha _2},$ $c\simeq c_0\phi ^{\alpha
_3}, $ $d\simeq d_0\phi ^{\alpha _4}$ in Eqs.(7) and balancing the different
terms, we obtain the following results:

\begin{equation}
\begin{array}{c}
\alpha _1=\alpha _2=\alpha _3=\alpha _4=-1, \\  
\\ 
a_0b_0+c_0d_0=-\frac 12. 
\end{array}
\end{equation}
\\For finding the powers at which the arbitrary functions can enter into the
series, we substitute the expressions,

\begin{equation}
\begin{array}{c}
a=a_0\phi ^{-1}+a_j\phi ^{j-1}, 
b=b_0\phi ^{-1}+b_j\phi ^{j-1} \\  \\ 
c=c_0\phi ^{-1}+c_j\phi ^{j-1},d=d_0\phi ^{-1}+d_j\phi ^{j-1} 
\end{array}
\end{equation}
\\into Eqs. (7), and compairing the lowest-order terms we obtain a system of
four linear algebraic equations in ($a_j$ $,$ $b_j$ $,$ $c_j$ $,$ $d_j$ ).
In matrix form it may be conveniently written as

\begin{equation}
\left[ A\left( j\right) \right] \left[ X\right] =0 
\end{equation}
\\where $\left[ X\right] $ $=(a_j$ $,$ $b_j$ $,$ $c_j$ $,$ $d_j$ $)^T$\\and\\
$\left[ A(j)\right] =\left[ 
\begin{array}{cccc}
\begin{array}{c}
\begin{array}{c}
B \\  
\end{array}
\\  
\end{array}
& 
\begin{array}{c}
\begin{array}{c}
-6 
a_0^2 \\ +3a_0^2(j-2) 
\end{array}
\\  
\end{array}
& 
\begin{array}{c}
\begin{array}{c}
-6 
a_0d_0 \\ +3a_0d_0(j-2) 
\end{array}
\\  
\end{array}
& 
\begin{array}{c}
\begin{array}{c}
-6 
a_0c_0 \\ +3a_0c_0(j-2) 
\end{array}
\\  
\end{array}
\\ 
\begin{array}{c}
\begin{array}{c}
-6 
b_0^2 \\ +3b_0^2(j-2) 
\end{array}
\\  
\end{array}
& 
\begin{array}{c}
B \\  
\end{array}
& 
\begin{array}{c}
\begin{array}{c}
-6 
b_0d_0 \\ +3b_0d_0(j-2) 
\end{array}
\\  
\end{array}
& 
\begin{array}{c}
\begin{array}{c}
-6 
b_0c_0 \\ +3b_0c_0(j-2) 
\end{array}
\\  
\end{array}
\\ 
\begin{array}{c}
\begin{array}{c}
-6 
b_0c_0 \\ +3b_0c_0(j-2) 
\end{array}
\\  
\end{array}
& 
\begin{array}{c}
\begin{array}{c}
-6 
a_0c_0 \\ +3a_0c_0(j-2) 
\end{array}
\\  
\end{array}
& 
\begin{array}{c}
C \\  
\end{array}
& 
\begin{array}{c}
\begin{array}{c}
-6 
c_0^2 \\ +3c_0^2(j-2) 
\end{array}
\\  
\end{array}
\\ 
\begin{array}{c}
-6 
b_0d_0 \\ +3b_0d_0(j-2) 
\end{array}
& 
\begin{array}{c}
-6 
a_0d_0 \\ +3a_0d_0(j-2) 
\end{array}
& 
\begin{array}{c}
-6 
d_0^2 \\ +3d_0^2(j-2) 
\end{array}
& C 
\end{array}
\right] $\\where

$B=(j-1)(j-2)(j-3)-3(j-1)-6a_0b_0+3(j-2)+3$ and\\$
C=(j-1)(j-2)(j-3)-3(j-1)-6c_0d_0+3(j-2)+3$\\To have a non-trivial solution
for $a_j$ $,$ $b_j$ $,$ $c_j$ and $d_j,$we demand that

\begin{equation}
\det A(j)=0 
\end{equation}
\\On solving Eqs. (11), we get the resonance values as $
j=-1,0,0,0,2,2,2,3,4,4,4,4.$\\The resonance at j = -1 corresponds to the
arbitrariness of $\psi \left( z,t\right) .$ On equating the coefficients of $
\psi ^{-4}$, we get a unique equation defining $a_0$, $b_{0,}c_0$ and $d_0$
which is given by

\begin{equation}
a_0b_0+c_0d_0=-\frac 12. 
\end{equation}
This shows that any three of the four functions $a_0$, $b_{0,}c_0$ and $d_0$
are arbitrary which corresponds to $j=0,0,0.$

Proceeding further and equating the coefficients of $\left( \psi ^{-3},\psi
^{-3},\psi ^{-3},\psi ^{-3}\right) ,$ we obtain

\begin{equation}
\begin{array}{c}
a_1= 
\frac{a_0}{3i\lambda }, \\  \\ 
b_1=- 
\frac{b_0}{3i\lambda }, \\  \\ 
c_1= 
\frac{c_0}{3i\lambda }, \\  \\ 
d_1=-\frac{d_0}{3i\lambda }. 
\end{array}
\end{equation}

On the other hand, the coefficients of $\left( \psi ^{-2},\psi ^{-2},\psi
^{-2},\psi ^{-2}\right) $ in Eqs. (7) reduce to a single equation

\begin{equation}
b_0a_2+a_0b_2+d_0c_2+c_0d_2=\frac{
\psi _t}{6\lambda } 
\end{equation}
so that three of the four functions $a_2$, $b_{2,}$ $c_2$ and $d_2$ are
arbitrary which corresponds to $j=2,2,2.$

Similarly from the powers of $\left( \psi ^{-1},\psi ^{-1},\psi ^{-1},\psi
^{-1}\right) $and $\left( \psi ^{-0},\psi ^{-0},\psi ^{-0},\psi ^{-0}\right)
,$ we find that Eqs. (4) admit the sufficient number of arbitrary functions
and hence Eqs. (7) possess the Painlev\'e property and hence they are
expected to be integrable.

\begin{center}
{\bf III. HIROTA BILINEARIZATION}
\end{center}

Hirota's bilinear method [20] is one of the most direct and elegant methods
available to generate multi-soliton solutions of nonlinear partial
differential equations. To avoid mathematical complexities, it is rather
convenient to transform Eqs.[4] to a simpler form, so that we may able
obtain multi-soliton solutions. We make the following transformations to
convert CHNLS to complex modified K-dV (cmK-dV) equation:

\begin{equation}
\begin{array}{c}
u\left( z,t\right) =Q_1\left( Z,T\right) Exp\left[ -i\left( \frac Z{3\lambda
}-\frac T{27\lambda ^2}\right) \right] , 
v\left( z,t\right) =Q_2\left( Z,T\right) Exp\left[ -i\left( \frac
Z{3\lambda }-\frac T{27\lambda ^2}\right) \right] \\  \\ 
t=T,Z=z+\frac t{3\lambda } 
\end{array}
\end{equation}
\\Using the above transformations in Eqs.[4], the resultant cmK-dV equation
is obtained in the form

\begin{equation}
\begin{array}{c}
Q_{1T}-\lambda \left[ Q_{1ZZZ}+6\left( \left| Q_1\right| ^2+\left|
Q_2\right| ^2\right) Q_{1Z}+3Q_1\left( \left| Q_1\right| ^2+\left|
Q_2\right| ^2\right) _Z\right] =0 \\  
\\ 
Q_{2T}-\lambda \left[ Q_{2ZZZ}+6\left( \left| Q_1\right| ^2+\left|
Q_2\right| ^2\right) Q_{2Z}+3Q_2\left( \left| Q_1\right| ^2+\left|
Q_2\right| ^2\right) _Z\right] =0 
\end{array}
\end{equation}
In order to construct Hirota's bilinear form, we consider Hirota bilinear
transformations in the form

\begin{equation}
Q_1=\frac GF{\hspace{1cm}}Q_2=\frac HF 
\end{equation}
\\where $G\left( Z,T\right) $ and $H\left( Z,T\right) $ are complex
functions and $F\left( Z,T\right) $ is a real function. Now using the
transformations (17), (16) can be rewritten as

\begin{equation}
\begin{array}{c}
F^2\left[ \left( D_T-\lambda D_Z^3\right) \left( G.F\right) \right] -\lambda 
[\{-3D_z^2\left( F.F\right) D_Z+12\left( \left| G\right|
^2+\left| H\right| ^2\right) D_Z\}\left( G.F\right)
 + \\   \\ 3GFD_Z\left( G.G^{*} \right) +3H^{*}FD_Z\left( H.G\right) -3
HFD_Z\left( G.H^{*}\right) ]=0   \\ \\
F^2\left[ \left( D_T-\lambda D_Z^3\right) \left( H.F\right) \right] -\lambda
[\{-3D_z^2\left( F.F\right) D_Z+12\left( \left| G\right|
^2+\left| H\right| ^2\right) D_Z\}\left( H.F\right) +   \\ \\
3HFD_Z\left( H.H^{*}\right) -3G^{*}
FD_Z\left( H.G\right) +3GFD_Z\left( H.G^{*}\right) =0 
\end{array}
\end{equation}
\\where the Hirota bilinear operators $D_z$ and $D_t$ are defined as

\begin{center}
\begin{equation}
D_Z^mD_T^nG\left( Z,T\right) F\left( Z^{\prime },T^{\prime
}\right) =\left( \frac \partial {\partial Z}-\frac \partial
{\partial Z^{^{\prime }}}\right) ^m\left( \frac \partial {\partial 
T}-\frac \partial {\partial T^{^{\prime }}}\right) ^nG\left(
Z,T\right) F\left( Z^{\prime },T^{\prime }\right) \mid _{Z=
Z^{^{\prime }},T=T^{\prime }} 
\end{equation}
\newpage\ {\bf IV. EXACT \ SOLITON\ SOLUTIONS}
\end{center}

Eqs.(18) can be decoupled into a set of bilinear equations as:

\begin{equation}
\begin{array}{c}
\left( D_T-\lambda D_Z^3\right) \left( G.F\right) =0 
{\hspace{.5cm}}\left( D_T-\lambda D_Z^3\right) \left( H.F\right) =0 \\  
\\ 
D_Z^2\left( F.F\right) =4\left( \left| G\right| ^2+\left| H\right| ^2\right)
\\  
\\ 
D_Z\left( G.G^{*}\right) =0,D_Z\left( H.H^{*}\right) =0,
D_Z\left( G.H^{*}\right) =0,D_Z\left( H.G^{*}\right) =0,
D_Z\left( G.H\right) =0 
\end{array}
\end{equation}
\\ In order to obtain soliton solutions, we are applying a perturbative
technique by writing the variables $F,$ $G,$ $H$ as a series in an arbitrary
parameter $\varepsilon $:

\begin{equation}
F=1+\varepsilon ^2f_2+\varepsilon ^4f_4+\cdot \cdot \cdot \cdot
\cdot \cdot \cdot ,{\hspace{.5cm}}G=\varepsilon g_1+\varepsilon
^3g_3+\varepsilon ^5g_5+\cdot \cdot \cdot \cdot \cdot \cdot ,{
\hspace{.5cm}}H=\varepsilon h_1+\varepsilon ^3h_3+\varepsilon
^5h_5+\cdot \cdot \cdot \cdot \cdot \cdot 
\end{equation}

\begin{center}
{\bf A. One-soliton solutions}
\end{center}

For one-soliton solution (1SS), we assume solutions in a series form in 
$\varepsilon $ such that

\begin{equation}
F=1+\varepsilon ^2f_2{\hspace{.5cm}}G=\varepsilon g_1
{\hspace{.5cm}}H=\varepsilon h_1 
\end{equation}
\\Substituting Eq.(22) in Eqs.(20) and then collecting coefficents of terms
with same powers in $\varepsilon $, we obtain:\\$\varepsilon :$\hspace{.5cm} 
\begin{equation}
\left( D_T-\lambda D_Z^3\right) \left( g_1.1\right) =0\hspace{.5cm}
\left( D_T-\lambda D_Z^3\right) \left( h_1.1\right) =0 
\end{equation}
\\$\varepsilon ^2:$\hspace{1cm} 
\begin{equation}
\begin{array}{c}
D_Z^2\left( 1. 
f_2+f_2.1\right) =4\left( g_1.g_1^{*}+h_1.h_1^{*}\right) \\  \\ 
D_Z(g_1.g_1^{*})=0,D_Z(h_1.h_1^{*})=0,D_Z(g_1.h_1^{*})=0,
D_Z(h_1.g_1^{*})=0,D_Z(g_1.h_1)=0 
\end{array}
\end{equation}
\\$\varepsilon ^3:$\hspace{1cm} 
\begin{equation}
\left( D_T-\lambda D_Z^3\right) \left( g_1.f_2\right) =0
\hspace{.5cm}\left( D_T-\lambda D_Z^3\right) \left( h_1.f_2\right) 
\end{equation}
\\$\varepsilon ^4:$\hspace{1cm} 
\begin{equation}
D_Z^2\left( f_2.f_2\right) 
\end{equation}
\\One can easily check that the solution, which is consistent with the
system (23-26), is

\begin{equation}
\begin{array}{c}
g_1=\cos \phi 
\exp \left( \eta \right) \hspace{.5cm}h_1=\sin \phi \exp
\left( \eta \right) \\  \\ 
f_2=\left( \frac 1{2k^2}\right) \exp \left( 2\eta \right) 
\end{array}
\end{equation}
\\where 
\begin{equation}
\eta =kZ+\lambda k^3T 
\end{equation}
\\and $\phi $ and $k$ are real constants. Using Eqs. (27) and (28) in (22)
and then in (17), after absorbing $\varepsilon $ the one-soliton solution
can easily be worked out to be

\begin{equation}
\begin{array}{c}
Q_1=\left( \frac k{
\sqrt{2}}\right) \cos \phi { sech}\left( kZ+\lambda 
k^3T+\eta _0\right) \\  \\ 
Q_2=\left( \frac k{\sqrt{2}}\right) \sin \phi { sech}\left( k
Z+\lambda k^3T+\eta _0\right) 
\end{array}
\end{equation}
\\where $\eta _0$ is a real constant. Using Eqs.(15), one-soliton solutions
of Eqs.(4) are found to be

\begin{equation}
\begin{array}{c}
u=\left( \frac k{
\sqrt{2}}\right) \cos \phi \exp \left[ -i\left( \frac Z{3\lambda
}-\frac T{27\lambda ^2}\right) \right] { sech}\left( kZ+\lambda 
k^3T+\eta _0\right) \\  \\ 
v=\left( \frac k{\sqrt{2}}\right) \sin \phi \exp \left[ -i\left(
\frac Z{3\lambda }-\frac T{27\lambda ^2}\right) \right]  sech\left( k
Z+\lambda k^3T+\eta _0\right) 
\end{array}
\end{equation}
\newpage\ 

\begin{center}
{\bf B. Two-soliton solutions}
\end{center}

The two-soliton solutions can be obtained by terminating the series
expansion for $F,$ $G,$ $H$ as:

\begin{equation}
F=1+\varepsilon ^2f_2+\varepsilon ^4f_4,{\hspace{.5cm}}
G=\varepsilon g_1+\varepsilon ^3g_2,{\hspace{.5cm}}
H=\varepsilon h_1+\varepsilon ^3h_3 
\end{equation}
\\and proceeding as before to obtain

\begin{equation}
\begin{array}{c}
g_1=\exp \left( \eta _1\right) + 
exp\left( \eta _2\right) {\hspace{.5cm}}h_1=ig_1 \\  \\ 
g_3=\left( k_2-k_1\right) ^2\left[ 
\frac{{exp}\left( 2\eta _1+\eta _2\right) }{4
k_1^2\left( k_1+k_2\right) ^2}+\frac{{exp}\left( \eta _1+2\eta _2\right) }{4k_2^2\left( k_1+k_2\right) ^2}\right] 
\hspace{.5cm}h_3=ig_3 \\  \\ 
f_2=4\left[ 
\frac{\exp \left( 2\eta _1\right) }{4k_1^2}+2\frac{\exp
\left( \eta _1+\eta _2\right) }{\left( k_1+k_2\right) ^2}+\frac{\exp \left( 2
\eta _2\right) }{4k_2^2}\right] \\  \\ 
f_4=\frac{4\left( k_2-k_1\right) ^4\exp \left( 2\eta _1+2
\eta _2\right) }{16k_1^2k_2^2\left(
k_1+k_2\right) ^4} 
\end{array}
\end{equation}
\\where

\begin{equation}
\eta _j=k_jZ+\lambda k_j^3T{\hspace{.5cm} }
j=1,2 
\end{equation}
\\Here $k_j$ is a real constant. Using Eqs.(32) and (33) in (31) and then in
(17), the two-soliton solutions of (16) are obtained. Using Eqs.(15), the
two-soliton solutions of Eqs.(4) are found to be:

\begin{equation}
\begin{array}{c}
u=\frac GF 
\exp \left[ -i\left( \frac Z{3\lambda }-\frac T{27\lambda ^2}\right)
\right] \\  \\ 
v=\frac HF\exp \left[ -i\left( \frac Z{3\lambda }-\frac T{27\lambda
^2}\right) \right] 
\end{array}
\end{equation}
\\Both 1SS and 2SS are in exact agreement with Eqs.(4).Two dimensional and
three dimensional plots of 2SS are given in figs.(1) and (2) respectively.
\newpage\ 

\begin{center}
{\bf V. LAX PAIR FOR CHNLS SYSTEM}
\end{center}

The linear eigen value problem associated with Eqs.(16) are [21,22]

\begin{equation}
\psi _Z=U\psi {\hspace{.5cm}}\psi _T=V\psi 
{\hspace{.5cm}}\psi =\left( \psi _1\psi _2\right) ^T 
\end{equation}
\\where

\begin{equation}
U=\left( 
\begin{array}{ccccc}
-i\chi & Q_1 & Q_1^{*} & Q_2 & Q_2^{*} \\ 
-Q_1^{*} & i\chi & 0 & 0 & 0 \\ 
-Q_1 & 0 & i\chi & 0 & 0 \\ 
-Q_2^{*} & 0 & 0 & i\chi & 0 \\ 
-Q_2 & 0 & 0 & 0 & i\chi 
\end{array}
\right) 
\end{equation}

\begin{equation}
\begin{array}{c}
V=\frac{-8i\lambda \chi ^3}5\left( 
\begin{array}{ccccc}
-4 & 0 & 0 & 0 & 0 \\ 
0 & 1 & 0 & 0 & 0 \\ 
0 & 0 & 1 & 0 & 0 \\ 
0 & 0 & 0 & 1 & 0 \\ 
0 & 0 & 0 & 0 & 1 
\end{array}
\right) -4\lambda \chi ^2\left( 
\begin{array}{ccccc}
0 & Q_1 & Q_1^{*} & Q_2 & Q_2^{*} \\ 
-Q_1^{*} & 0 & 0 & 0 & 0 \\ 
-Q_1 & 0 & 0 & 0 & 0 \\ 
-Q_2^{*} & 0 & 0 & 0 & 0 \\ 
-Q_2 & 0 & 0 & 0 & 0 
\end{array}
\right) \\  
\\ 
+2i\lambda \chi \left( 
\begin{array}{ccccc}
-2A & -Q_{1Z} & -Q_{1Z}^{*} & -Q_{2Z} & -Q_{2Z}^{*} \\ 
-Q_{1Z} & \left| Q_1\right| ^2 & (Q_1^{*})^2 & Q_1^{*}Q_2 & Q_1^{*}Q_2^{*}
\\ 
-Q_{1Z}^{*} & Q_1^2 & \left| Q_1\right| ^2 & Q_1Q_2 & Q_1Q_2^{*} \\ 
-Q_{2Z} & Q_1Q_2^{*} & Q_1^{*}Q_2^{*} & \left| Q_2\right| ^2 & (Q_2^{*})^2
\\ 
-Q_{2Z}^{*} & Q_1Q_2 & Q_1^{*}Q_2 & Q_2^2 & \left| Q_2\right| ^2 
\end{array}
\right) \\  
\\ 
+\lambda \left( 
\begin{array}{ccccc}
0 & a_{12} & a_{13} & a_{14} & a_{15} \\ 
a_{21} & a_{22} & 0 & a_{24} & a_{25} \\ 
a_{31} & 0 & a_{33} & a_{34} & a_{35} \\ 
a_{41} & a_{42} & a_{43} & a_{44} & 0 \\ 
a_{51} & a_{52} & a_{53} & 0 & a_{55} 
\end{array}
\right) 
\end{array}
\end{equation}
\\where\\$a_{12}=4$ $A$ $Q_1+Q_{1ZZ},$ $a_{13}=4$ $A$ $Q_1^{*}+Q_{1ZZ}^{*},$ 
$a_{14}=4$ $A$ $Q_2+Q_{2ZZ},$ \\$a_{15}=4$ $A$ $Q_2^{*}+Q_{2ZZ}^{*}$\\$
a_{21}=-4$ $A$ $Q_1^{*}-Q_{1ZZ}^{*},$ $a_{22}=Q_1Q_{1Z}^{*}-Q_1^{*}Q_{1Z},$ $
a_{24}=Q_2Q_{1Z}^{*}-Q_1^{*}Q_{2Z},$ $%
a_{25}=Q_2^{*}Q_{1Z}^{*}-Q_1^{*}Q_{2Z}^{*}$\\$a_{31}=-4$ $A$ $Q_1-Q_{1ZZ},$ $
a_{33}=Q_1^{*}Q_{1Z}-Q_1Q_{1Z}^{*},$ $a_{34}=Q_2Q_{1Z}-Q_1Q_{2Z},$ $
a_{35}=Q_2^{*}Q_{1Z}-Q_1Q_{2Z}^{*}$\\$a_{41}=-4$ $A$ $Q_2^{*}-Q_{2ZZ}^{*},$ $
a_{42}=Q_1Q_{2Z}^{*}-Q_2^{*}Q_{1Z},$ $%
a_{43}=Q_1^{*}Q_{2Z}^{*}-Q_2^{*}Q_{1Z}^{*},$ $%
a_{44}=Q_2Q_{2Z}^{*}-Q_2^{*}Q_{2Z}$\\$a_{51}=-4$ $A$ $Q_2-Q_{2ZZ},$ $%
a_{52}=Q_1Q_{2Z}-Q_2Q_{1Z},$ $a_{53}=Q_1^{*}Q_{2Z}-Q_2Q_{1Z}^{*},$ $%
a_{55}=Q_2^{*}Q_{2Z}-Q_2Q_{2Z}^{*}$\\with $A=\left| Q_1\right| ^2+\left|
Q_2\right| ^2.$

The compatibility condition $U_T-V_Z+\left[ U ,V\right] =0$ gives
rise to Eqs.(16). The construction of Lax pair confirms that Eqs.(16) and
thereby the CHNLS Eqs.(4) are indeed completely integrable.

\begin{center}
{\bf VI.CONCLUSION}
\end{center}

In this paper, we have considered a set of coupled NLS equations with
higher-order linear and nonlinear dispersion terms included. Then, by
choosing the parameters as in the case of the corresponding integrable
uncoupled case, we applied the Painlev\'e singularity structure analysis and
established that for this particular choice of parameters, Eqs.(4) possess
the Painlev\'e property. We have explictly obtained one-soliton and
two-soliton solutions for the integrable cases of CHNLS equations using
Hirota bilinearisation technique and solutions are plotted. We have also
constructed Lax pairs using AKNS formalism. Hence, with these results, we
have proved that the CHNLS equations which describe the wave propagation of
two fields in fibre systems with all higher-order effects such as TOD, Kerr
dispersion and stimulated Raman effect, will allow soliton-type propagation.
From the soliton solutions, one can obtain information about the shape,
width and intensity of the propagation pulse.{\it \ }

\begin{center}
{\bf ACKCOWLEDGEMENTS}
\end{center}

The authors are thankful to the referee for valuable comments and would like
to thank Prof. M. Lakshmanan and Dr. K. Porsezian for valuable discussions.
\ One of us (MNV) would like to thank CSIR, New Delhi for financial support
in the form of JRF. VCK acknowledges Associateship of IUCAA, Pune.

\begin{enumerate}
\item  Figure 1a : 3D profile of $\left| u(z,t)\right| $ for the two-soliton
solution of Eq. (34) with the parameter values k$_1$=0.034, k$_2$=0.04, $%
\lambda =0.005$

\item  Figure 1b : 2D profile of $\left| u(z,t)\right| $ for the two-soliton
solution of Eq. (34) with the parameter values k$_1$=2, k$_2$=3, $\lambda
=0.002$

\item  Figure 1c : Contour plot of $\left| u(z,t)\right| $ with respect to z
and t for the parameter values k$_1$=0.034, k$_2$=0.04, $\lambda =0.005$

\item  Figure 2a : 3D profile of $\left| v(z,t)\right| $ for the two-soliton
solution of Eq. (34) with the parameter values k$_1$=0.04, k$_2$=0.045, $%
\lambda =0.005$

\item  Figure 2b : 2D profile of $\left| v(z,t)\right| $ for the two-soliton
solution of Eq. (34) with the parameter values k$_1$=2, k$_2$=3, $\lambda
=0.005$

\item  Figure 2c : Contour plot of $\left| u(z,t)\right| $ with respect to z
and t for the parameter values k$_1$=0.04, k$_2$=0.045, $\lambda =0.005$
\end{enumerate}
\end{document}